\documentclass[a4paper,11pt]{article}

\usepackage[margin=2.5cm]{geometry}
\usepackage{mathptmx}
\usepackage{hyperref}
\hypersetup{bookmarksnumbered=true,
	pdftitle={Foundations of biology},
	pdfauthor={J.-L. Sikorav, A. Braslau, and A. Goldar},
	pdfkeywords={economy epistemology evolution history methodology pedagogy},
	}
\usepackage[nosort]{cite}
\usepackage{mdwlist}
\usepackage{multicol}

\renewenvironment{description*}{%
\begin{basedescript}{%
}%
}{%
\end{basedescript}%
}

\begin{document}

\noindent
{\Huge\bf Foundations of biology}
\\[1em]
\rule{\textwidth}{1bp}\\[5bp]
{\large J.-L. Sikorav$^1$, A. Braslau$^2$ and A. Goldar$^3$}\\
\rule{\textwidth}{1bp}
\\[1em]
{\it
$^1$DSM, Institut de Physique Th{\'e}orique, IPhT, CNRS, MPPU, URA2306, CEA/Saclay, F-91191 Gif-sur-Yvette, France.
\\
$^2$DSM, Service de Physique de l'{\'E}tat Condens{\'e}, SPEC, CNRS URA2464, CEA/Saclay, F-91191 Gif-sur-Yvette, France.
\\
$^3$DSV, iBiTec-S, Service de Biologie Int{\'e}grative et de G{\'e}n{\'e}tique Mol{\'e}culaire, CEA/Saclay, F-91191 Gif-sur-Yvette, France.
}

\begin{abstract}
\noindent
It is often stated that there are no laws in biology, where everything is contingent and could have been otherwise, being solely the result of historical accidents. Furthermore, the customary introduction of fundamental biological entities such as individual organisms, cells, genes, catalysts and motors remains largely descriptive; constructive approaches involving deductive reasoning appear, in comparison, almost absent. As a consequence, both the logical content and principles of biology need to be reconsidered.\par
The present article describes an inquiry into the foundations of biology. The foundations of biology are built in terms of elements, logic and principles, using both the language and the general methods employed in other disciplines. This approach assumes the existence of a certain unity of human knowledge that transcends discipline boundaries. Leibniz's principle of sufficient reason is revised through the introduction of the complementary concepts of symmetry and asymmetry and of necessity and contingency. This is used to explain how these four concepts are involved in the elaboration of theories or laws of nature. Four fundamental theories of biology are then identified: cell theory, Darwin's theory of natural selection, an informational theory of life (which includes Mendel's theory of inheritance) and a physico-chemical theory of life. Atomism and deductive reasoning are shown to enter into the elaboration of the concepts of natural selection, individual living organisms, cells and their reproduction, genes as well as catalysts and motors.\par
This work contributes to clarify the philosophical and logical structure of biology and its major theories. This should ultimately lead to a better understanding of the origin of life, of system and synthetic biology, and of artificial life.

\end{abstract}

\noindent
keywords: economy epistemology evolution history methodology pedagogy

\begin{multicols}{2}

\section{Introduction}

Our work investigating the foundations of biology began as a theoretical construction of the genetic material. We observed that the fundamental concept of a material basis to heredity is commonly introduced through a description of the DNA double helix, followed by another description of its replication process. \cite{Watson1953a,Watson1953c} Asking ``Why DNA is such as it is and not otherwise?'', we chose to pursue a complementary approach: Our focus was on a construction in the general deductive spirit of the work of Crane, \cite{Crane1950} establishing a minimal list of requirements that a biological device for information storage should possess. The genetic material emerged from this construction as a transient state in a succession of invariant processes of replication. Over the years, this work has been slowly mended through presentations to various audiences, in research seminars, meetings and courses (both at the undergraduate and graduate levels), and has benefited from the criticisms of many people. The approach is of pedagogical interest, as it improves our intuitive understanding of the structure and function of DNA. Encouraged by the positive features of this construction, we sought to better understand the nature of the rules upon which it was made.\par
The structure of DNA, when compared to that of the hypothetical genetic material as obtained theoretically, is --- in a qualified sense --- unique and ideal. This conclusion contrasts with a widely held opinion, according to which there are no laws in biology, where everything is contingent and could have been otherwise, being solely the result of historical accidents. It thus raises the broad question of the role of necessity in this discipline. Asking ``Why life is such as it is and not otherwise?'', we were inevitably induced to ponder on the foundations of biology and its connection with the problems of the origin of life, of artificial life and of synthetic biology. Finally, we came to compare this with the foundations of other disciplines.\par

\subsection{The foundations of knowledge}

The disciplines of logic, mathematics and physics have gone through foundational crises leading to a deeper understanding of their elements, logic and principles and their historical development. In mathematics, the concept of number (for example, real or infinite) has been revised, novel axioms as well as constructive approaches have been introduced and new measures defined (such as the modern theory of measure). In physics, the concepts of time and of space, of mass and of energy, of temperature, and of identity and of causality have been reconsidered and metrology has been completely renovated. Modern physics has identified many new elements (such as atoms as well as subatomic particles), and this has required the elaboration of new logical systems (in statistical physics and in quantum mechanics). The role of probability theory in the logic of science has grown considerably. Logic itself has been transformed by G{\"o}del's work, showing that certain propositions can neither be proved nor refuted, thus giving an unexpected role to contingency in this discipline, and radically transforming the concept of demonstration. Various types of logical systems have developed in the last century: finite or not, discrete or continuous, multivalued or fuzzy, temporal (bearing on past and future, on retrodiction and prediction) or describing other modalities. As a result, the largely illusory {\it a priori} character of logic has faded away, the process of differentiation into diverse systems making patent (using the wording of Quine) its naturalization. Lastly, the broad importance of the concept of symmetry has been grasped progressively in logic, in mathematics, in physics, and in the elaboration of theories or laws of nature, as further explained below.\par
From this logical point of view, the disciplines mentioned above appear to have common characteristics: the importance of their deductive content, as well as the key role played by elemental objects in the logical inferences employed. A certain ``elemental logic'', bearing on items that are irreducible and invariant, is at the heart of many constructive approaches. This resembles the logical atomism of Russell \cite{Russell1911} which has been a source for modern analytical philosophy.\par
Biology differs in several ways: It has not gone through a foundational crisis. Furthermore, it is often claimed that biology is unique, and that contingency reigns, providing the ultimate explanation for everything. Such statements both deny the existence of theories in biology and, more generally, of a unity of knowledge. The logical content, especially deductive, of biological knowledge remains uncertain. This raises the question of the role of atomism in biology, a question having many facets as well as ancient roots. The three great biological theories introduced in the nineteenth century, namely cell theory, Darwin's theory of natural selection, \cite{Darwin1859} and Mendel's theory of inheritance \cite{Mendel1866,Bateson1902} were formulated at a time when the atomic structure of matter was still being debated. This implies that their initial formulation must be amended in order to make room for an elemental logic. We note that Mendel's theory is often considered as a main root of atomism in biology, introducing implicitly the concept of particulate inheritance, as put forth by Fisher, \cite{Fisher1930} thus correcting the erroneous idea of blending inheritance. \cite{Rostand1950,deBeer1964} Yet Mendel's ``elements'', later called genes by Johannsen, \cite{Johannsen1909} are now known to be compound items rather than true units of biological information. Furthermore, modern physical atomism originates from other sources, namely the kinetic theory of gases and the theory of Brownian motion, and its general consequences for biological thought remain to be explored.

\subsection{Aim of the present work}

The goal of the present work is threefold: first to provide a succinct exposition of our work on the foundations of biology and of the methodology used, sketching the main arguments and major conclusions; secondly, to describe a research program; thirdly, to present a pedagogical project. A companion article describes theoretical constructions of the genetic material and of proteins using the results obtained here. \cite{Sikorav2014construction}\par
In the study of the foundations of biology, one must identify elemental phenomena of the discipline and describe the measures employed to understand them and the resulting logic. At the same time, the underlying principles must be disclosed, bringing into light philosophical considerations. Foundations are also to be studied in historical terms.\par
Working on all these aspects of the foundations of biology is an immense task that can not be completed in a few years by a small group of researchers. It constitutes a research program, similar to the mathematical Erlangen program of Felix Klein \cite{Klein1872} seeking to incorporate group theory into the study of geometry.\par
Finally, we also think of foundations in terms of what we believe should be taught first, in terms of elementary or basic phenomena, concepts and theories. We have found over the years that the foundations of biology can be to a large extent taught at the undergraduate level (and even before). The present work should, therefore, be viewed also as a pedagogical project.\par

\section{Results and discussion}

Biology is a scientific discipline based on phenomena derived from observations or experiments, on concepts elaborated from them, and on further generalizations derived from these concepts through theories or laws of nature. We investigate its foundations using the language and the methods employed in the investigation of disciplines outside of biology. Our approach is based on a fundamental and general belief in the existence of a unity of human knowledge. This assumption is, nevertheless, constantly held as a working hypothesis susceptible of refutation.\par

\subsection{Revising the principle of sufficient reason}

The question ``Why is DNA such as it is and not otherwise?'' can be addressed, following Leibniz, using the philosophical principle of sufficient reason. This states that we can always provide an answer to the following two questions: ``Why does something exist rather than nothing?'', and, if it exists, ``Why is it as such and not otherwise?''. The broad importance of this principle is well known: As remarked by Enriques, \cite{Enriques1909a} it constitutes a postulate of the intelligibility of reality as well as a requirement for the elaboration of models, and provides rules for scientific constructions. At the same time, its practical use is not so easily grasped. First, this principle is constrained by the observation that, most often, the reasons shall remain unknown to us. This limitation illustrates the general finding that fundamental principles never come alone, but always appear in complementary pairs, in a dialectic manner.\par
Underlying the principle of sufficient reason are the concepts of necessity and contingency and of symmetry and asymmetry. To make the use of the principle of sufficient reason easier, we have broken it into four sub-principles, defining four philosophical attitudes towards these dual couples of antithetic concepts, illustrated in Table~\ref{tab:fourattitudes}. One can associate, for instance, symmetry with contingency through the classical point of view according to which contingency arises from ignorance, a lack of information. Conversely, one can associate asymmetry with necessity by observing that a phenomenon, to occur, requires the absence of certain symmetry elements, in other words, the presence of necessary asymmetries; Thirdly, symmetries can be taken as necessary, focusing on simplicity, economy and invariances. Fourthly, asymmetries can be taken as contingent, focusing on imagination and invention. These four attitudes can be used to explain how the concepts of necessity and contingency and of symmetry and asymmetry enter in the formation of knowledge, in particular, in the principle of sufficient reason and in the elaboration of theories or laws of nature.\par
\nocite{Simon1981}
\nocite{Laplace1820,Keynes1921}
\nocite{Gibbs1902,Shannon1948,Jaynes1957}
\nocite{Weyl1952,Weyl1949,Wigner1967,Lee1981}
\nocite{Wiener1923}
\nocite{Simon1981}
\nocite{Curie1894}
\nocite{Carnot1824}
\nocite{Perrin1909}
The construction of a theory describing a phenomenon (or a set of phenomena) consists of two steps: The first is the elaboration of appropriate measures aiming at its study. Secondly, from these measures can then be identified the invariants of the phenomenon which include both necessary asymmetries and compatible symmetries. Phenomenal asymmetries can be deemed necessary according to Curie's asymmetry principle. \cite{Curie1894} In a complementary manner, the set of all the symmetries compatible with a phenomenon defines its symmetry group. Such symmetries can be viewed philosophically as being either contingent or necessary. The necessity of symmetry in mathematics is found in a principle of symmetry, stated by Weyl, \cite{Weyl1952} according to which any mathematical object must be characterized in terms of the set of symmetries, called its automorphism group, leaving it invariant. This represents an extension of Klein's program for geometry to the whole of mathematics and then to the natural sciences. The method of construction of a theory that we follow, following Lautman, \cite{Lautman1946} therefore aims both to extract the necessary asymmetries from a phenomenon and to incorporate as many symmetry elements as possible in the symmetry group which describes it. In this approach, we do not construct something that actually exists, but rather an ideal, Platonic structure. We follow a similar approach in the theoretical construction of a compound item. A necessary item is unique and, if it is compound, can be assembled from elemental components using the rules of construction given above. In contrast, an item that is not necessary is called contingent as it could be otherwise; a contingent item cannot be constructed, but only described.\par
The claim that everything in biology is contingent raises a general philosophical problem as to the origin of necessity (outside of biology). We view necessity as having its sources both in logic and symmetry considerations. In formal logic, necessity is associated with deductive reasoning (being called an apodictic necessity by Kant). \cite{Kant1883} Extending this idea to the natural sciences, we state that the necessary character of an inference made in any scientific discipline is related to its content in deductive reasoning. The inference leading, for instance, to the establishment of a difference between two objects using a measure of sufficient accuracy possesses such a deductive character. It allows us to hold as necessary the conclusion of non-identity. This can explain, in part, the origin of the necessity found in Curie's asymmetry principle. Another major source of necessity arises from the attitude associating it with symmetry, at the heart of modern physics. Making symmetry necessary has the consequence to confer a character of certainty to the generalization of induction.\par

\subsection{Biology and physical atomism}

The introduction of atomism in biology is required to deduce the existence of individual living organisms (meaning literally, that cannot be divided) and to construct plausible schemes for a process of reproduction. The basic logical tool here is derived from Fermat's principle of infinite (or indefinite) descent (which, in his terms, can be used both in a negative and in an affirmative manner). \cite{Fermat1894} The Fermatian inference further rules-out reproduction processes based on a constant reduction of the size of (nestled) living organisms (as in the theory of the ``homunculus''). This was understood by Buffon, \cite{Buffon1749} who, however, could not draw his argumentation to a clear conclusion as the size and reality of atoms themselves were unknown in his days. Another consequence of atomism is that reproduction cannot proceed unbounded under finite resources, pointing to the necessary existence of some sort of competition or struggle for life, a statement that is at the basis of the theory of natural selection. One can thus conclude that physical atomism plays a central role in the construction of biological concepts and theories. In the following section, we shall see how it enters in the four fundamental theories of life: the theory of natural selection, cell theory, the informational theory of life and the physico-chemical theory of life.\par

\subsection{Four fundamental theories of biology}

\renewcommand{\thesubsubsection}{\arabic{subsubsection}.}

\subsubsection{The theory of natural selection}

The process of reproduction has two aspects: it is a kind of multiplication, a nonlinear event, and also a certain transmission of hereditary information, implying a memory system. Any system involving a memory requires its persistence through time. This requirement is absolute and cannot tolerate even the slightest interruption, thus relying implicitly on a principle of continuity (also known as Leibniz's law of continuity). Biology which relies on reproduction of individuals is, therefore, based on a pair of seemingly contradictory yet actually complementary principles: of continuity and discontinuity. The simultaneous presence of both continuity and discontinuity is found in all sciences, as observed by d'Arcy Thompson, \cite{DArcyThompson1942} among others.\par
The principle of discontinuity (or atomicity) is central to elemental logic. Continuity, on the other hand, is also present in the theory of natural selection because living organisms are made of a very large number of finite, minute constituents that we perceive as a continuum. This constitutes a second aspect of the principle of continuity illustrating a general principle of fine division. Indeed, many phenomena that we study are complex, involving a large number of elementary components usually interacting in a non-trivial manner. Elements are often few in type yet each is present in large number, an expression of a principle of plenty, complementary to the principle of economy or parsimony. A statistical approach to such systems is appropriate, and their description can be made in a purely continuous manner even though a knowledge of the underlying atomic features is essential for their full understanding.\par
Darwin observed that reproduction is not an exact invariance, but an approximate symmetry that generates heritable, in general minute, differences. \cite{Darwin1859} He then generalized by induction this empirical finding to all living organisms. The underlying current explanation is that both the number of hereditary traits and the informational content of each one (in terms of number of bits) are very large. Consequently, this makes possible the existence of a very large number of variations, each slight enough to be treated as infinitesimal, compatible with the use of a continuous description of the reproduction process. Any individual organism must, accordingly, be thought of as a complex network of interactions between numerous hereditary traits and the environment. (Our present understanding describes in greater details this network in terms of interaction between genes and their products: RNA, proteins and other molecules.)\par
A close look at the logic behind Darwin's theory of natural selection leads to the identification of several underlying, complementary principles and of three main types of logical inferences:
\begin{description*}
\item [inductive,] such as the statement that the reproduction of individuals is not an exact invariance, but an approximate symmetry that generates heritable differences,
\item [deductive,] proving the existence of a process called natural selection leading to the retention of certain living organisms and to the elimination of others, is a necessary consequence of the imperfection of reproduction and of the finiteness of available resources and
\item [Bayesian, {\rm \cite{Laplace1820}}] leading (when augmented by concepts elaborated in cell theory, see below) to the parsimonious conclusion of the existence of a unique common ancestor for all living organisms.
\end{description*}
The inductive as well as the key deductive components of Darwin's theory were identified long ago by Julian Huxley \cite{Huxley1942} and by Leigh van Valen. \cite{vanValen1976} More recent works, such as Mayr's book on the history of biology or several articles written in 2009 (150 years after the publication of the Origin), discuss the logic of this theory following Huxley in terms of five major inferences, \cite{Mayr1982,Ayala2009,Gregory2009} yet of unspecified status (deductive, inductive or otherwise).\par
Natural selection defines a second law of interaction between living organisms (reproduction being the first). This invariant process provides a constant means of dispersal, acting on all forms of life following the last common ancestor. The existence of this common ancestor, when compared with the immense variety of present living organisms, brings to the fore the action of a principle of divergence underlying natural selection. Natural selection explains in a parsimonious manner the immense variety of living organisms in their totality. Indeed, of all the theories conceived by man to understand the universe, the theory of natural selection stands amongst the simplest in hypotheses and the richest in phenomena.\par
Finally, natural selection operates through a constant search for extremes, both maxima and minima, over all parameters at its disposition. We observe the results of this search of extremes for instance in terms of size, of largeness and minuteness, at the level of entire organisms (with elephants and whales, for example, at one end and prokaryotic single-cell organisms at the other). At the cellular level, the eggs of birds are single cells of macroscopic size; similarly, the largest neurons are of the size of an entire animal and can be several meters long. At the molecular level, natural selection leads to the formation of the giant macromolecules of meter-long chromosomal DNA but also to a constantly increasing repertoire of small molecules such as secondary metabolites. At the metabolic level, this search for extremes leads to states of maximum as well as minimum dissipation of energy. The overall measure of natural selection, introduced by Fisher, is called fitness and has the dimension of the reciprocal of time. \cite{Fisher1930} The relative fitness of living organisms is to be understood as an expression a biological principle of fine division.\par

\subsubsection{Cell theory}

Our examination of cell theory also exploits the principle of atomicity, and we construct cell theory starting from the requirements of natural selection: reproduction, imperfect transmission of hereditary information and constant operation of selective pressure. The feasibility of this constructive approach is again based on the principle that excludes a division {\it ad infinitum} of living organisms. Concomitantly, the theory of natural selection tells us that cells, being complex entities, are themselves assembled from a very large number of elements and can reasonably be described as continuous objects. Our approach is, therefore, semi-continuous. Elaborated in this manner, cell theory is both deductive and inductive and completes the nineteenth century view (of Schleiden, Schwann and others) which remained purely inductive.\par
The simplest cellular shape is spherical. The process of reproduction of a cell of minimal size requires cellular growth prior to a binary fission; this combined process of growth followed by division, that we call cell gemination, makes the spherical symmetry of cells only approximate, as observed d'Arcy Thompson \cite{DArcyThompson1942} and others. Binary fission creates a singular point of contact between two daughter cells at the moment of separation. This point becomes a new pole, leading to the deduction that all cells are polarized, both in space and in time, in accordance with the principle of fine division. (Similar statements can be found in the literature, for instance in the work of Taddei and coworkers \cite{Stewart2005}.) We emphasize here the logic behind cell theory.\par
Cell theory constructs both states and processes. It defines different types of cells: the simplest are those involved in ordinary uniparental or haploid reproduction, the more complex (which are germ cells or gametes in addition to somatic cells) are involved in biparental or diploid reproduction. Cell theory states that haploid generation preceded diploid generation. Furthermore, the common ancestor of all living organisms was unicellular. It was perhaps not a isolated entity, but a set of cells freely sharing their genomic content that later evolved into distinct organisms. \cite{Woese1998}\par
Cell theory defines the state of latent life \cite{Bernard1878}, also called cryptobiosis \cite{Keilin1959}, in which living organisms can survive for extended periods of time as closed or isolated thermodynamic systems in a reversible, dormant stage. Cryptobiosis is both necessary and universal as a potentiality at the cellular level. Thus, at any time, a cell is either living, in a cryptobiotic state, or dead. Cell death is a necessity dictated by natural selection. Yet, every cell living today is connected to the common ancestor by a lineage of cells, all of which have escaped death. The continuity of cellular life thus implies that cell death, although necessary, is not universal. Cell theory also constructs the process of programmed cell death, necessary and universal as a potentiality in living organisms today. Both programmed cell death and cryptobiosis are necessary to ensure the continuity of life across most adverse conditions. In unicellular organisms, programmed cell death is often an intermediate process in a pathway leading to cryptobiosis.\par

\subsubsection{The informational theory of life}

The theory of biological inheritance or informational theory of life envisions living organisms, their structures, their functions and their interactions, in terms of information, stored or transmitted, emitted or received, and involved in regulation or control, in homeostasis \cite{Cannon1929} as well as adaptation. Biology can be seen as a branch of information and communication sciences, and of cybernetics. \cite{Wiener1961,Ashby1960} Biological information constitutes a fundamental asymmetry of life (comparable with molecular chirality, often itself called the asymmetry of life). Information is a physical concept and is associated with an energy cost. \cite{Szilard1929} The informational theory of life, as such, includes the study of heredity or genetics, and we focus here on this aspect. The theory of inheritance has its roots, in part, in the work of Mendel in the nineteenth century. \cite{Mendel1866,Bateson1902} The logical basis of this theory, however, was developed mostly starting in the second half of the twentieth century.\par
The work of von~Neumann on self-reproducing automata offers a proof of existence of such a logic. \cite{vonNeumann1951,vonNeumann1966} The architecture of his six-component automaton is described in Table~\ref{tab:vonNeumann} and compared to the structure of living organisms. The self-reproducing automaton is built from two fundamental, distinct components: a set of instructions and an aggregate of smaller automata. This distinction is required in order to comply with the temporal logic of the process of reproduction. It agrees with the necessary existence of two types of informational biopolymers and thus of a genetic code relating their sequences. \cite{Crick1958} Wigner's ``no-cloning'' theorem of quantum mechanics, furthermore, rules out the possibility of an exact reproduction. \cite{Wigner1961,Wigner1967} Shannon's theory of communication makes possible gradual changes through redundancy and error correction. \cite{Shannon1948} The discontinuous or particulate nature of inheritance, logically derived from cell theory and from information theory, ensures a law of conservation: the retention of the variation, at the basis of the process of natural selection.\par
One can use these ideas to construct a complete theory of inheritance, first for the simplest living organisms reproducing through uniparental, haploid generation, then for the more complex organisms reproducing through biparental, diploid generation. Assuming by simplicity that gametes possess a haploid system of inheritance, one can deduce first the existence of a system of diploid inheritance in zygotes and somatic cells and then establish rigorously Mendel's law of inheritance.\par
The concepts of genetic linkage and genetic recombination can be introduced to be used later in a construction of the genetic material. The seven pairs of characters investigated by Mendel are transmitted independently of one another. Later experiments with sweet peas \cite{Bateson1905} have shown that for certain pairs of genes a preferred co-transmission can be observed, a phenomenon called genetic linkage, leading to the elaboration of the concept of linkage group. In complex living organisms such as metazoan, the genes usually clustered into several linkage groups. In simpler prokaryotic unicellular organisms, such as the bacterium {\it Escherichia coli}, there can exist a single linkage group. \cite{Lederberg1947} This latter discovery relied in fact on the exploitation of the newly discovered phenomenon of genetic recombination in bacterium. \cite{Lederberg1946} Indeed, investigations of genetic linkage revealed the plasticity of linkage groups and led to the elaboration of the concept of genetic recombination according to which genes do not occupy fixed positions, but can occasionally move either within a given linkage group or to a different linkage group. This has been observed both in eukaryotes and in prokaryotes. We shall formulate a generalization of these observations stating that recombination is a universal process in living organisms. The process of genetic recombination points to the existence of cut and paste tools operating on the genetic material within or between linkage groups.\par

\subsubsection{The physico-chemical theory of life}

The physico-chemical theory of life seeks to understand the process of natural selection using the concepts of physics and chemistry. This theory includes molecular biology (and extends into the infra-molecular level, as both nuclear and electronic properties have to be considered, as well as into the supra-molecular level, indeed up to the macroscopic level). We focus here on the elemental level. This theory tries to explain the high rates, high yields and high specificity of biochemical reactions and how efficient transport processes operate. It is through the study of the biochemistry, physical chemistry and biophysics of nucleic acids that we came to study the general questions raised here. Indeed, we described methods coupling chemical reactions and phase transitions that increase the rates of nucleic-acid hybridization and cyclization by many thousand-fold. These investigations fostered our initial interest in heterogeneous biochemistry and for a principle of fine division underlying biochemical fitness. This principle of fine division manifests itself through complementariness, of heterogeneity (such as phase heterogeneity), anisotropy (such as molecular chirality \cite{Pasteur1848}) and of fine division of time. It can be understood as an extension of the Carnot principle for heat engines \cite{Carnot1824} to other sources of energy.\par
Chemical catalysis (a concept introduced by Berzelius and developed by Ostwald, Fischer, Pauling, among others) also involves the principle of fine division as it operates through a tight, supramolecular complementariness between the catalyst and a transition state conformation occurring in the chemical reaction. The most general definition of a catalyst is that of an invariant transforming a virtual process allowed by the laws of thermodynamics, but which will not necessarily occur, into a phenomenon observable within a finite time. Both the biological process of reproduction and von~Neumann's self-replicating automaton are instances of (auto)catalysis. Catalysis defines a cycle, conveniently drawn in the Krebs representation. \cite{Krebs1946} In a Carnot cycle, for instance, the motor itself is a catalyst and is not consumed; In another example, a carbon nucleus serves as the catalyst in the cyclic process of energy production in stars. \cite{Bethe1939a} Chemical catalysts contribute to fitness by increasing the rates of biochemical reactions and are thus necessary biological entities. Catalysis (as originally perceived by Berzelius \cite{Berzelius1836}) brings together the fields of engineering, chemistry and biology, illustrating a unity of knowledge combining the sciences of nature with the sciences of the artificial.\par
The physico-chemical theory of life investigates the relation between motion and life. The identification of movement with life and stillness with death is ancient. The observation of the phenomenon of Brownian motion in microscopic studies by Robert Brown \cite{Brown1828,Brown1829} renewed the question of this relationship. Unger and Dujardin distinguished Brownian motion from active cell motion, killing cells either by heating or through the addition of poisons. \cite{Unger1832,Harris1999,Dujardin1835} Further questions are raised today by the study of (supramolecular) biological motors.\par
Mitchell's chemiosmotic theory endeavors to unite transport and metabolism into a vectorial metabolism and introduces Curie's asymmetry principle in biochemistry. \cite{Mitchell1967} Osmoenzymes (in the language of Mitchell \cite{Mitchell1979a,Mitchell1991}), now called biological motors, perform various types of mechanical work, improving, in particular, the transport of compounds beyond that permitted by Brownian motion, being, therefore, just as catalysts, universal and necessary biological entities.\par
The physico-chemical theory of natural selection explains the extraordinary stability of the components of living organisms in the state of cryptobiosis. The necessary existence of specific macromolecules, heteropolymers, whether catalysts, motors or informational devices for the storage of hereditary information, can then be established deductively. Such approaches can explain the unique character of essential biological structures and functions, leading, in particular, to a clearer, intuitive understanding of the structures of proteins and nucleic acids, described in a separate article.\par

\section{Conclusions and Perspectives}

Our approach adapts the methods used in the investigation of other disciplines to biology, and our emphasis is on the methods themselves as much as on results. Underlying this approach is a conviction of the existence of a unity of knowledge, which finds its main manifestation in a science of research (Leibniz {\it Ars~inveniendi} \cite{Couturat1901} or Peirce's {\it Economy of Research} \cite{Peirce1976}), describing methods shared by all disciplines. Yet, the tree of knowledge also grows through the many disciplinary branches, the existence of which can be explained in terms of an increased efficiency associated with a division of intellectual labor at the basis of economical employment of thought. We find here again fundamental principles coming as complementary pairs: a principle of diversity expressing a pluralism coupled with a principle of unity, as well as the idea that efficiency is associated with a certain fine division, whether in the process of natural selection or in the economy of knowledge.\par
The belief in the unity of knowledge, common to students of methods embodied by Leibniz and Peirce, is shared by a large number of thinkers (for instance Schr{\"o}dinger, \cite{Schroedinger1944} E.~O. Wilson \cite{Wilson1998}, Eigen \cite{Eigen2013}, and others), even in an age of increasing specialization. One can mention here the current interest in transdisciplinarity and convergence.\par
The unity of human knowledge is also found in the foundations of biology, which are observed here to rely on elemental phenomena, on pairs of complementary principles (of continuity and discontinuity or atomicity, of symmetry and asymmetry, of necessity and contingency, of parsimony and plenty, ...), and on logic, in the same manner as do other sciences of nature. We have seen, for instance, how the foundations of biology ultimately rely through atomism on an elemental logic. Maxwell's conclusion on the need to incorporate atomism in biological thought remains as timely today as it was almost a century and a half ago. \cite{Maxwell1875} The foundations of biology provide this discipline both with elements, laws of interaction and theories, making possible a deductive formulation of the concept of natural selection and then of cells, genes, catalysts and motors. They will explain in a plausible manner the necessary and unique character of essential biological structures and functions.\par

\paragraph{Acknowledgements}
The present work summarizes an ongoing research program initiated a few years ago. Previous versions of it have been presented to various audiences: Carg{\`e}se summer school on DNA and Chromosomes, in 2004, 2006, and 2009; Gent Fantom Research School on Symmetries and symmetry violation in 2004 and 2007; Lectures on the Foundations of Molecular Biology, Evry in 2004; Lectures on Elements of Biology, Ecole Polytechnique F{\'e}d{\'e}rale de Lausanne in 2008; Lectures on Foundations of Biology, University Pierre and Marie Curie, Paris 2008-2013; 74th Cold Spring Harbor Symposium ``Evolution --- The Molecular Landscape'' in 2009 and Meeting ``From Base Pair to Body Plan'', Cold Spring Harbor Laboratory in 2013. We have greatly benefited from the comments of their participants. We wish to thank for discussions and/or comments the manuscript Roger Balian, Sydney Brenner, George Church, Gregory Chaitin, Albert Libchaber, Marie-Claude Marsolier-Kergoat, Theo Odijk, Monica Olvera de la Cruz and Edouard Y{\'e}ramian.\par

\end{multicols}


\clearpage

\begin{table}[!ht]
\caption{Four attitudes towards necessity and contingency, symmetry and asymmetry}
\begin{tabular} {@{}|r|c|c|@{}}
\hline
 & Contingency & Necessity \\
\hline
Symmetry &
\parbox{.36\linewidth}{\raggedright\strut
{\it ``Incompletism''}\\
Principle of bounded rationality \cite{Simon1981}\\
Principle of indifference \cite{Laplace1820,Keynes1921}\\
Principle of maximum entropy \cite{Gibbs1902,Shannon1948,Jaynes1957}
} &
\parbox{.43\linewidth}{\raggedright\strut
{\it ``Symmetrism''}\\
Principle of simplicity or economy\\
Principle of symmetry\\
Conservation laws, selection rules, symmetries as ``non-observables'' \cite{Weyl1952,Weyl1949,Wigner1967,Lee1981}
} \\
\hline
Asymmetry &
\parbox{.36\linewidth}{\raggedright\strut
{\it ``Inventism''}\\
Gedanken experiments of physics\\
Mathematical Brownian motion \cite{Wiener1923}\\
Sciences of the artificial \cite{Simon1981}
} &
\parbox{.43\linewidth}{\raggedright\strut
{\it ``Phenomenism''}\\
Curie's principle of asymmetry \cite{Curie1894}\\
Heat engines \cite{Carnot1824}\\
Brownian motion and molecular reality \cite{Perrin1909}
} \\
\hline
\end{tabular}
\\[.5em]
\begin{minipage}{\textwidth}
\small
The four attitude may be viewed as forming a modified modal square of opposition (yet we stress their complementary --- and not their contradictory --- nature).\par
\renewcommand{\descriptionlabel}[1]{\hspace{\labelsep}\emph{#1}}
\begin{description}
\item[Incompletism] associates symmetry with asymmetry, in a principle of ``non-sufficient'' reason or indifference. Both in physics and in information theory, ignorance, viewed as a lack of information, can prevent the establishment of the uniqueness of a phenomenon. This is treated through a maximization of the relevant (Gibbs-Boltzmann or Shannon) entropies.\par
\item[Phenomenism] associates asymmetry with necessity.
The presence of necessary asymmetries for the occurrence of a phenomenon constitutes Curie's asymmetry principle (sometimes erroneously called Curie's symmetry principle). For instance, a macroscopic heat engine requires:
(1) a transfer between two reservoirs differing in temperature (a space heterogeneity),
(2) an absence of time reversibility (time orientation) and
(3) an oriented heat flow from the hot to the cold reservoir.
This Carnot principle can be viewed as a manifestation of a principle of division of the simplest type relating division with efficiency. A Curie analysis of the phenomenon of the Brownian motion of a colloidal particle in a fluid similarly reveals two fundamental spatial and temporal asymmetries. It establishes the granularity of the surrounding fluid, its discrete, atomic (or molecular) structure (and, therefore, the disruption of the symmetry of scale invariance); it also reveals the constant motion of this fluid (in agreement with the kinetic theory of heat).\par
\item[Symmetrism] asserts the necessity of symmetry at the foundations of laws of nature. It arises first through the principle of simplicity or economy: when several laws can describe the same phenomenon, we must retain the one that is the simplest, most parsimonious, devoid of redundancy. Simplicity can be here defined and measured through algorithmic information theory (as developed by Solomonoff, Kolmogorov and Chaitin). The information content of a phenomenon is formally given by the size of the smallest algorithm characterizing it. In symmetrism, phenomena are not described through necessary asymmetries, but instead through a group of compatible symmetries, either continuous or discrete, and the focus is on conservation laws or selection rules. Symmetries are called ``non-observables'', and, accordingly, asymmetries become associated observables as in phenomenism.\par
\item [Inventism] associates asymmetry with contingency. Here, one decides to suppress some of the necessary asymmetries of phenomenon, treating them as contingent. It releases imagination and contributes to creative thinking at large. Gedanken (thought) experiments of physics lead to the prediction of phenomena, thus to conjectural asymmetries. It can also be found in all disciplines that are not constrained by natural phenomena such mathematics or the sciences of the artificial. Mathematical Brownian motion (also called a Wiener process) can be obtained as the continuous limit of a discrete random walk on a lattice and possesses the property of unbounded scale and time invariance. Yet both scale and time invariance are incompatible with the phenomenon of Brownian motion and thus mathematical Brownian motion does not describe physical reality.\par
\end{description}
\end{minipage}
\label{tab:fourattitudes}
\end{table}

\begin{table}[!ht]
\caption{The self-replicating automaton of von~Neumann}
\begin{tabular} {@{}|p{.115\textwidth}|p{.38\textwidth}|p{.435\textwidth}|@{}}
\hline
Component & Function in the automaton & Structural counterpart in living organisms \\
\hline
$\mathrm{I}_{\mathrm{D}}$ & The instruction describing automaton~D & DNA \\
A & Constructing automaton & Transcription and translation machinery \\
B & Copying automaton & DNA replication machinery \\
C & Controlling automaton & Regulation of replication and gene expression \\
D & The aggregate $\mathrm{A}+\mathrm{B}+\mathrm{C}$ & A cell without its genetic material \\
E & The aggregate $\mathrm{A}+\mathrm{B}+\mathrm{C}+\mathrm{I}_{\mathrm{D}}$ & The cell, the simplest living organism \\
\hline
\end{tabular}
\\[.5em]
\begin{minipage}{\textwidth}
The automaton of von~Neumann consists of five automata and one instruction. This instruction $\mathrm{I}_{\mathrm{D}}$ is itself an aggregate of simple parts and acts as the tape in a Turing computing automaton. The automaton~B makes a copy of $\mathrm{I}_{\mathrm{D}}$: ``the copying mechanism~B performs the fundamental act of reproduction, the duplication of the genetic material, which is clearly the fundamental operation in the multiplication of living cells.'' \cite{vonNeumann1951} When the constructing automaton~A is furnished with the instruction $\mathrm{I}_{\mathrm{D}}$, the controlling automaton ``C will first cause A to construct the automaton which is described by this instruction $\mathrm{I}_{\mathrm{D}}$. Next C will cause B to copy the instruction $\mathrm{I}_{\mathrm{D}}$ referred to above, and insert the copy into the automaton referred to above, which has just been constructed by A. Finally, C will separate this construction from the system $\mathrm{A}+\mathrm{B}+\mathrm{C}$ and `turn it loose' as an independent entity.'' \cite{vonNeumann1951} In order to function, the aggregate $\mathrm{D}=\mathrm{A}+\mathrm{B}+\mathrm{C}$ must be furnished with the instruction $\mathrm{I}_{\mathrm{D}}$ describing this very automaton~D, thus forming the self-reproducing automaton~E. ``E is clearly self-reproductive. Note that no vicious circle is involved. The decisive step occurs in E, when the instruction $\mathrm{I}_{\mathrm{D}}$, describing D, is constructed and attached to D. When the construction (the copying) of $\mathrm{I}_{\mathrm{D}}$ called for, D exists already, and it is in no wise modified by the construction of $\mathrm{I}_{\mathrm{D}}$. $\mathrm{I}_{\mathrm{D}}$ is simply added to form E. Thus there is a definite chronological and logical order in which D and $\mathrm{I}_{\mathrm{D}}$ have to be formed, and the process is legitimate and proper according to the rules of logic.'' \cite{vonNeumann1951}
\end{minipage}
\label{tab:vonNeumann}
\end{table}



\end{document}